\documentclass[aps,prb,twocolumn,showpacs]{revtex4}

\usepackage[english]{babel}
\usepackage{graphicx, epsfig}
\usepackage{amscd}
\usepackage{amsmath,amsfonts,amssymb}
\usepackage{graphicx}

\usepackage{amsmath}

\begin{document}

\title{\bf
Quantum discord in spin systems with dipole-dipole interaction
}

\author{Elena I.\,Kuznetsova}
\email{kuznets@icp.ac.ru} 
\author{M. A. Yurishchev}

\affiliation{Institute of Problems of Chemical Physics of Russian Academy of Sciences,
Chernogolovka, 142432, Moscow Region, RUSSIA}

\date{}
\begin{abstract}
The behavior of total purely quantum correlation (discord)
in dimers consisting of dipolar-coupled spins 1/2 is studied.
We found that the discord $Q=0$ at the absolute zero temperature.
With increasing the temperature $T$, at first the quantum correlations in
the system  increase, smoothly reach the maximum, and then
turn again into zero according to the asymptotic law $T^{-2}$.
It is also shown that in absence of external magnetic field $B$,
the classical correlation $C$ at $T\to0$ is, vice versa, maximal.
Our calculations predict that in crystalline gypsum
${\rm CaSO_4\cdot2H_2O}$ the value of natural ($B=0$) quantum
discord between nuclear spins of hydrogen atoms is maximal
at the temperature of 0.644~$\mu$K, and for
1,2-dichloroethane ${\rm H_2ClC-CH_2Cl}$ the discord achieves
the largest value at $T=0.517~\mu$K.
In both cases, the discord equals $Q\approx0.083$~bit/dimer
what is $8.3\%$ of its upper limit in two-qubit systems.
We estimate also that for gypsum at room temperature
$Q\sim10^{-18}$~bit/dimer,
and for 1,2-dichloroethane at $T=90$~K the discord is 
$Q\sim10^{-17}$~bit per a dimer.
\end{abstract}

\pacs{03.65.Ud, 03.67.Mn, 75.10.Jm, 75.50.Xx}
\maketitle



\section{Introduction}
\label{sec:Intro}

The notion {\em quantum discord\/} has been introduced in the quantum
information theory by Zurek \cite{Z00}.
Discord was regarded as ``a measure of a violation of classicality of a
joint state of two quantum subsystems'' \cite{Z00}.
However later its initial definition undergone some changes. 

In 2001, Henderson and Vedral and then independent of them Ollivier
and Zurek in their papers \cite{HV01,OZ02} (see also \cite{V03,Z03})
performed analysis of every possible correlations $I$ in a bipartite 
system and suggested the ways to extract from them, on the
one hand, the purely classical part $C$ and, on the other hand,
only the quantum contribution $Q$.
The quantum excess of correlations, $Q=I-C$, has been called
``discord'' --- in the modern understanding of this term \cite{OZ02}.

The authors of above papers established also that the quantum
correlation can be non-zero even in separable (but mixed) states.
In other words, quantum correlations are not exhausted by
entanglement ($E$).
Entanglement, which can relate the different parts of a
system even when there are no interactions between these parts
(the Einstein-Podolsky-Rosen effect),
is only a special kind of quantum correlations.
Since the 80--90s of past century the entanglement was considered
as a fundamental resource for quantum information processing,
teleportation, cryptography, metrology, and other tasks in quantum
technology \cite{NCh01,VK02,V05,AFOV08,HHHH09}. 

It is remarkable that quantum discord can also lead to a speedup
over classical computation and lead even without containing much
entanglement \cite{DSC08,LBAW08}.
This important property of discord evoked extremely great interest
to the new kind of correlations.
One discovered also that discord can detect the quantum phase
transitions \cite{Dil08,Sar09}.
Moreover, it has been shown that in contrast to entanglement and
thermodynamical quantities, the discord makes it possible to catch
the approach of quantum phase transitions even at finite temperatures  
\cite{WTRR10}. 
Other important features of discord have been also noted.
By this, a surprising fact turned out:
``Almost all quantum states have nonclassical correlations''
\cite{FACCA10}.
Achieved up to now results on the theory and applications of quantum
discord are given in the recent reviews \cite{CMS11,MBCPV11}.

The goal of this paper is to study the behavior of discord in spin
systems with dipolar couplings.
There is a large number of substances magnetic interactions
in which have the dipole-dipole character, and exchange and indirect
ones are weak enough.
By this, the spins both electrons and nuclears can serve as
elementary magnetic moments. 
The class of dipolar magnets with electron spins includes, for
example, the Tutton salts, alums \cite{Kh50}, and numerous salts
of rare earth elements \cite{JM91}.
The typical temperatures at which the effects of their dipole
interactions show themselves lie in the millikelvin region.
However the spin-lattice interactions in electron paramagnets are
strong that leads to short relaxation times too.
Nuclear spins, of course, have an indubitable asset:
the spin-lattice relaxation times for them can achieve minutes and
hours.
Using the available NMR data for the two classical examples,
gypsum ${\rm CaSO_4\cdot2H_2O}$ \cite{P48} and 1,2-dichloroethane
${\rm H_2ClC-CH_2Cl}$ \cite{GKPP49} (see also, e.g.,
\cite{A61,G80}) which contain the sufficiently isolated pairs of
dipolar-coupled nuclear hydrogen spins, we estimate the discord
between these spins. 

In following sections of the paper we give definitions for the
different correlations, formulate the model, calculate the classical
and quantum information correlations in it, estimate the discord for
materials with spin-nuclear dimers, and, lastly, briefly summarize
the results obtained.

\section{Classical and quantum correlations}
\label{sec:defQ}

In statistical theory, a degree of relationship (correlation) between
two random variables $x$ and $y$ with the joint probability
distribution function $p(x,y)$ is often measured by covariations
or by Pearson's correlation coefficient 
\begin{equation}
   \label{eq:k}
   R=\frac{\overline{(x-\overline{x})(y-\overline{y})}}
   {\sqrt{D_1}\sqrt{D_2}} . 
\end{equation}
Here the bar denotes the average over the probability distribution
and $D_1=\overline{(x-\overline{x})^2}$ and 
$D_2=\overline{(y-\overline{y})^2}$ are dispersions respectively
for $x$ and $y$.
Notice that in the mathematical statistics other types of correlation
coefficients are used also.
One should emphasize that the condition $R=0$ does not imply,
generally speaking, the independence of random variables, i.~e., that 
$p(x,y)=p_1(x)p_2(y)$ \cite{KK68,Ch82}.

In the classical information theory (e.~g., \cite{M76,CT91}),
one uses the notion of mutual information 
\begin{equation}
   \label{eq:I}
   I(X:Y) = H(X) + H(Y) - H(X,Y) \ge 0
\end{equation}
between two objects $X$ Й $Y$.
Here $H(X)$, $H(Y)$, and $H(X,Y)$ are the Shannon entropies
\begin{eqnarray}
   \label{eq:HXY}
   &&H(X)= -\sum_x p_1(x)\,{\rm log}\,p_1(x),\nonumber\\
  && H(Y) = -\sum_y p_2(y)\,{\rm log}\,p_2(y),
   \nonumber\\
   &&H(X,Y) = -\sum_{x,y} p(x,y)\,{\rm log}\,p(x,y),
\end{eqnarray}
where $p_1(x)=\sum_y p(x,y)$, $p_2(y)=\sum_x p(x,y)$.
(A choose of logarithm base defines the information measure
unities: bits, nats, dits, hartley.)
It is remarkable that now the equality $I=0$ is a necessary and
sufficient condition for the independence of $X$ and $Y$.
This property allows to use the mutual information as a measure
of information correlation between the systems $X$ and $Y$
\cite{KK68}.

Taking into account the Bayes rule one can rewrite the right part
of Eq.~(\ref{eq:I}) in Shannon's nonsymmetric form  
\begin{equation}
   \label{eq:I1}
   I^\prime (X:Y) = H(X) - H(X|Y) ,
\end{equation}
where
\begin{equation}
   \label{eq:HX1Y}
   H(X|Y) = H(X,Y) - H(Y)
\end{equation}
is the condition entropy.
In the classical case, $I^\prime\equiv I$.

In the quantum information theory \cite{NCh01,H10}, the equation 
(\ref{eq:I}) is replaced by the new definition
\begin{equation}
   \label{eq:Iq}
   I(A:B) = S(\rho_A) + S(\rho_B) - S(\rho_{AB}) ,
\end{equation}
which serves as a measure of quantum mutual information between the
two subsystems $A$ and $B$ composing together the joint system
$AB=A\cup B$.
In Eq.~(\ref{eq:Iq}), $\rho_{AB}$ is the density matrix
of the joint system $AB$, $\rho_A$ and $\rho_B$ are the reduced
density matrices, respectively, for subsystems $A$ and $B$, and
$S(\rho)$ ($\rho=\{\rho_A, \rho_B, \rho _{AB}\}$) represents the
von Neumann entropy
\begin{equation}
   \label{eq:S}
   S(\rho) = -{\rm Tr}\,\rho\,{\rm log}\,\rho .
\end{equation}
It is important that $I=0$ is the necessary and sufficient condition
for the factorization 
$\rho_{AB}=\rho_A\otimes\rho_B$, what means, of course, the absolute
independence (non-correlativety) of $A$ and $B$ in the product state.
Therefore, in the quantum information theory, one takes the mutual
information to measure the total (both classical and quantum)
correlations between the two subsystems of bipartite quantum system.

On the other hand, measurements performed on one system, in general,
influence on the quantum state of another system (see, for instance,
\cite{H03}).
Postulating that the total classical part of correlations is the
{\em maximal\/} amount of information about one subsystem, say $A$,
that can be extracted by performing a measurement on the other
subsystem $B$, Henderson and Vedral \cite{HV01} suggested to take
as a measure of classical correlation the quantity 
\begin{equation}
   \label{eq:Ca}
   C(\rho_{AB}) = \max_{\{B_i\}}\{S(\rho_A) - \sum_ip_iS(\rho_A^i)\} .
\end{equation}
Here $\{B_i\}$ is a complete set of measurements on the subsystem
$B$,
\begin{equation}
   \label{eq:rho-ai}
   \rho_A^i={\rm Tr}_B(B_i\rho_{AB}B_i^+)/
   {\rm Tr}_{AB}(B_i\rho_{AB}B_i^+)
\end{equation}
is the remaining state of $A$ after obtaining the outcome $i$ on
$B$, and 
\begin{equation}
   \label{eq:p-i}
   p_i={\rm Tr}_{AB}(B_i\rho_{AB}B_i^+)
\end{equation}
is the probability to detect the result $i$.
 
Ollivier and Zurek \cite{OZ02}, on the contrary, focused their
attention on an extraction of quantum correlations.
Further analysis of the measurements led to the generalization of expression
(\ref{eq:I1}) for the quantum case,
\begin{equation}
   \label{eq:I1a}
   I^\prime(A:B) = S(\rho_A) - \sum_ip_iS(\rho_A^i) . 
\end{equation}
(It is obvious that the right hand of this equality is a non-optimized
classical correlation of Henderson and Vedral.)
In the paper \cite{OZ02}, the {\em minimal\/} difference   
$I-I^\prime\equiv Q$ has been identified with an amount
of quantum correlation and has been called the quantum discord ---
a measure of the quantum excess of correlations, a measure of the
quantumness of correlations.
Taking into account that
\begin{equation}
   \label{eq:ICQ}
   I = C + Q ,
\end{equation}
we see the equivalentness of results of Henderson-Vedral and
Ollivier-Zurek.

Quantum discord displays a number of properties (see, e.~g., the
reviews \cite{CMS11,MBCPV11}). 
We note between them the following ones.
For a pure state, discord coincides with the entanglement $E$.
In mixed states, the quantum correlation (discord) can present even
in that case when the entanglement is absent.
Quantum discord $Q\geq0$.
Discord is limited from above by the entropy of one subsystem,
$Q\leq S(\rho_{A(B)})$.
In particular, if the system is two-qubit and the logarithm base
for entropy equals two then $Q\leq1$.

\section{Hamiltonian and density matrix}
\label{sec:H-rho}

Consider a system consisting of two identical particles with
the spins 1/2 which couple between themselves by the magnetic
dipole-dipole interaction.
Let moreover, the external homogeneous magnetic field with induction
${\bf B}$ was applied to the system.
Then the Hamiltonian of a model can be written as (see, e.~g.,
\cite{NCh01})
\begin{equation}
   \label{eq:Hv}
   {\cal H} = {\cal H}_{dd} + {\cal H}_Z ,
\end{equation}
where the dipolar part is
\begin{equation}
   \label{eq:Hdd}
   {\cal H}_{dd} = \frac{\mu_0}{4\pi}\frac{\gamma^2\hbar^2}{4r^3}
   [{\bf\sigma_1}{\bf\sigma_2} 
   - 3({\bf n}\cdot{\bf\sigma}_1)({\bf n}\cdot{\bf\sigma}_2)]
\end{equation}
and the Zeeman energy equals
\begin{equation}
   \label{eq:HZ}
   {\cal H}_Z = -{1\over2}\gamma\hbar({\bf\sigma}_1
   + {\bf\sigma}_2){\bf B} . 
\end{equation}
In these equations, $\mu_0$ is the magnetic constant
(magnetic permeability of free space),
$\gamma$ is the gyromagnetic ratio,
${\bf\sigma}_{1,2}$ are the vectors of Pauli matrices at the sites
1 and 2,
$r$ is the distance between the spins in a dimer,
${\bf n}$ is the unit vector in the direction from one spin to
the other, and
$\bf B$ is the vector of magnetic field induction.
The dipole-dipole interaction reflects the exact law (in that sense
that it does not contain the fitting parameters),
the interaction is sharply anisotropic and long-acting (in contrast,
say, to the exchange interaction).
 
In the spherical coordinates when 
${\bf B}=(0,0,B)$ and ${\bf n}=(\sin\theta, 0, \cos\theta)$,
the Hamiltonian (\ref{eq:Hv}) -- (\ref{eq:HZ}) takes the form
\begin{eqnarray}
   \label{eq:Hs}
   &&{\cal H} = \frac{\mu_0}{4\pi}\frac{\gamma^2\hbar^2}{4r^3}
   [{\bf\sigma_1}{\bf\sigma_2} 
   - 3({\bf\sigma}_1^x\sin\theta + {\bf\sigma}_1^z\sin\theta) \nonumber\\ &&
   \times({\bf\sigma}_2^x\sin\theta + 
   {\bf\sigma}_2^y\sin\theta) ]  
   - {1\over2}\gamma\hbar B({\bf\sigma}_1^z + {\bf\sigma}_2^z). \nonumber\\ &&
\end{eqnarray}
It has been shown \cite{FMS11} that when the polar angle
$\theta={\pi\over2}$
(the external field is applied perpendicularly to the direction
of dimer longitudinal axis the entanglement in the system is maximal. 
On the contrary, when
$\theta=0$ or $\pi$ (the longitudinal dimer axis is parallel to the
external field) the entanglement between spins is absent 
(below we prove this strongly/exactly).

Because the special interest is to discover and investigate the
quantum correlations without entanglement, we will consider from
this point the case $\theta=0$.
By this, the Hamiltonian (\ref{eq:Hs}) takes the form 
\begin{equation}
   \label{eq:Hxxz}
   {\cal H} = {1\over2}D(\sigma_1^x\sigma_2^x 
   + \sigma_1^y\sigma_2^y 
   + \Delta\sigma_1^z\sigma_2^z) 
   - {1\over2}h(\sigma_1^z + \sigma_2^z) ,
\end{equation}
where the dipolar coupling constant equals
\begin{equation}
   \label{eq:D}
   D = \frac{\mu_0}{4\pi}\frac{\gamma^2\hbar^2}{2r^3} 
\end{equation}
and the normalized external field is
\begin{equation}
   \label{eq:h}
   h = \gamma\hbar B .
\end{equation}
In dipole-dipole coupled dimer, the anisotropy parameter is
$\Delta=-2$.
However, below we will, in some cases for the sake of generality,
extend the values of $\Delta$.
But all graphical and numerical material in our paper is presented
for $\Delta=-2$.

The Hamiltonian (\ref{eq:Hxxz}) corresponds to the XXZ model in Z
field.
In the matrix forn, it is given as
\begin{equation}
   \label{eq:Hxxz1}
   {\cal H}=\left(
      \begin{array}{cccc}
      \frac{\Delta}{2}D-h&&&\\
      &-\frac{\Delta}{2}D&D&\\
      &D&-\frac{\Delta}{2}D&\\
      &&&\frac{\Delta}{2}D+h
      \end{array}
   \right) .
\end{equation}
The $2\times2$ subblock presented here is a centrosymmetric matrix
which under the orthogonal transformation
\begin{equation}
   \label{eq:O}
   O=\frac{1}{\sqrt2}\left(
      \begin{array}{rr}
      1&1\\
      1&-1
      \end{array}
   \right) 
\end{equation}
undegoes in diagonal form.

The energy spectrum of the Hamiltonian (\ref{eq:Hxxz1}) with condition
$\Delta=-2$ consists of two independent on external field levels
\begin{equation}
   \label{eq:E12}
   E_1 = 2D,\qquad E_2=0
\end{equation}
and two levels
\begin{equation}
   \label{eq:E34}
   E_{3,4} = -D\pm h .
\end{equation}
Because $D>0$, the ground state energy is 
\begin{equation}
   \label{eq:E0}
   E_0 = -D + |h| .
\end{equation}
In absence of external field, the ground state is two-fold 
degenerate.

We will consider the dimer in a termal equilibrium state. 
In this case, its density matrix $\rho\equiv\rho_{AB}$ has the Gibbs
form
\begin{equation}
   \label{eq:rho}
   \rho={1\over Z}\exp(-\beta{\cal H}),
\end{equation}
where $\beta=1/k_BT$, $k_B$ is Boltzmann's constant, and
$Z$ is the partition function,
\begin{equation}
   \label{eq:Zd}
   Z={\rm Tr}\exp(-\beta{\cal H}).
\end{equation}
Performing necessary calculations we find that in the original
(standard) basis 
$|00\rangle$, $|01\rangle$, $|10\rangle$ Й $|11\rangle$ 
the density matrix has the structure
\begin{equation}
   \label{eq:rho-abvd}
   \rho=\left(
      \begin{array}{rrrr}
      a&&&\\
      &b&v&\\
      &v&b&\\
      &&&d
      \end{array}
   \right),
\end{equation}
where
\begin{equation}
   \label{eq:abdv}
   \begin{array}{c}
   a={1\over Z}\exp[-\beta({\Delta\over2}D - h)], \\ \\
   d={1\over Z}\exp[-\beta(\frac{\Delta}{2}D + h)], \\ \\
   b={1\over Z}\exp(\beta\frac{\Delta}{2}D){\rm ch}\,\beta D,\\ \\
   v=-{1\over Z}\exp(\beta\frac{\Delta}{2}D){\rm sh}\,\beta D,
   \end{array}
\end{equation}
and partition function is
\begin{equation}
   \label{eq:Z}
   Z=2({\rm ch}\,\beta D + e^{-\beta D\Delta}{\rm ch}\,\beta h)
   e^{\beta D\Delta/2}.
\end{equation}
Expressions (\ref{eq:abdv}) satisfy the condition
\begin{equation}
   \label{eq:ad2b}
   a + d + 2b = 1 ,
\end{equation}
which provides the normalization ${\rm Tr}\rho =1$.

Expanding the density matrix (\ref{eq:rho-abvd}) into powers of
Pauli matrices we obtain it in the (normal) Bloch form 
\begin{eqnarray}
   \label{eq:rhoB}
    &&\rho={1\over4}[1 + (a-d)(\sigma_1^z +\sigma_2^z) 
   + 2v(\sigma_1^x\sigma_2^x + \sigma_1^y\sigma_2^y) \nonumber \\
  && \quad + (1-4b)\sigma_1^z\sigma_2^z] .
\end{eqnarray}
Expansion coefficients are the unary and binary correlators,
\begin{equation}
   \label{eq:m}
   m\equiv\langle\sigma_1^z\rangle=\langle\sigma_1^z\rangle=a-d
   = {2\over Z}\,e^{-\beta D\Delta/2}{\rm sh}\,\beta h,
\end{equation}
\begin{equation}
   \label{eq:Gzz}
   G_\parallel\equiv\langle\sigma_1^z\sigma_2^z\rangle=1-4b
   = 1 - {4\over Z}\,e^{\beta D\Delta/2}{\rm ch}\,\beta D,
\end{equation}
\begin{equation}
   \label{eq:Gperp}
   G_\perp\equiv\langle\sigma_1^x\sigma_2^x\rangle
   = \langle\sigma_1^y\sigma_2^y\rangle = 2v
   = - {2\over Z}\,e^{\beta D\Delta/2}{\rm sh}\,\beta D .
\end{equation}
Brackets denote the trace operation for the expression in
brackets with density operator,
$\langle\cdot\rangle={\rm Tr}(\cdot\rho)$.
The coefficient $a-d=m$ in the expansion (\ref{eq:rhoB}) is equal
to the $z$ components (projections) of the Bloch vectors for the
reduced density matrices of subsystems $A$ and $B$. 
Moreover, the quantities $m$, $G_\parallel$, and $G_\perp$ have
a physical sense, correspondingly, as the normalized magnetization,
longitudinal and transverse components of correlation matrix.
Relations  
\begin{eqnarray}
   \label{eq:abvd}
   &&a={1\over4}\,(1+2m+G_\parallel),\qquad
   b={1\over4}\,(1-G_\parallel),
   \nonumber\\
   &&v={1\over2}\,G_\perp,\qquad
   d={1\over4}\,(1-2m+G_\parallel)
\end{eqnarray}
give the expressions for the matrix elements of the density
operator (\ref{eq:rho-abvd}) through the system correlators.

From Eq.~(\ref{eq:m})--(\ref{eq:Gperp}) we find the high temperature
behavior for the magnetization and correlation functions,
\begin{equation}
   \label{eq:m-as}
   m(T, h)={1\over2}\,\frac{h}{k_BT}
   - \frac{\Delta}{4}\,\frac{h}{D}\left(\frac{D}{k_BT}\right)^{\!2}
   + O(1/T^3) ,
\end{equation}
\begin{eqnarray}
   \label{eq:Gzz-as}
   &&G_\parallel(T, h)=-\frac{\Delta}{2}\,\frac{D}{k_BT}
   -{1\over4}\left[1
   -\left(\frac{h}{D}\right)^{\!2}\right]\!\left(\frac{D}{k_BT}\right)^{\!2} \nonumber \\ &&
  +O(1/T^3) ,
\end{eqnarray}
\begin{eqnarray}
   \label{eq:Gperp-as}
   && G_\perp(T, h)=-{1\over2}\,\frac{D}{k_BT}
   -\frac{\Delta}{4}\left(\frac{D}{k_BT}\right)^{\!2}
   \nonumber \\ && +{1\over8}\Big[{1\over3} 
   +\left(\frac{h}{D}\right)^{\!2}\Big]\!\left(\frac{D}{k_BT}\right)^{\!3}
   +O\left(\frac{1}{T^4}\right) . \nonumber \\ &&
\end{eqnarray}
Thus, when $T\to\infty$, the main contributions to the
correlators tend to zero according to the law $1/T$.
By this, a weak external field does not exercise an influence
on the $G_\parallel$ and $G_\perp$.
Moreover, the leading term in the expansion of the
transverse correlator $G_\perp$ does not depend on the
anisotropy $\Delta$. 

At lower temperature, the correlation functions for the dipolar
dimer ($D>0$, $\Delta=-2$) in absence of external field behave as
\begin{equation}
   \label{eq:Gzz-ash0}
   G_\parallel\vert_{T\to0}\approx
   1-\exp(-\frac{D}{k_BT}) ,
\end{equation}
\begin{equation}
   \label{eq:Gperp-ash0}
   G_\perp\vert_{T\to0}\approx
   -{1\over2}\exp(-\frac{D}{k_BT}). 
\end{equation}
If the field $h\neq0$, the additional statistical weights at the
exponents arise (due to a change of ground state of the system;
see below),
\begin{equation}
   \label{eq:Gzz-ash}
   G_\parallel\vert_{T\to0}\approx
   1-2\exp(-\frac{D}{k_BT}) ,
\end{equation}
\begin{equation}
   \label{eq:Gperp-ash}
   G_\perp\vert_{T\to0}\approx
   -\exp(-\frac{D}{k_BT}). 
\end{equation}
Thus, the lower temperature behavior of correlators is described
by a function of the form $e^{-1/x}$.

\section{Quantum entanglement}
\label{sec:E}

The entanglement through the relation
\begin{eqnarray}
   \label{eq:EC}
  && E=
   -\frac{1+\sqrt{1-{\tilde C}^2}}{2}\,
   {\rm log}_2\left(\frac{1+\sqrt{1-{\tilde C}^2}}{2}\right) \nonumber \\
  && -\frac{1-\sqrt{1-{\tilde C}^2}}{2}\, 
   {\rm log}_2\left(\frac{1-\sqrt{1-{\tilde C}^2}}{2}\right)
\end{eqnarray}
is expressed via the concurrence $\tilde C$ \cite{HW97,W98}.

In the case of density matrix having the block-diagonal form 
\begin{equation}
   \label{eq:rhobd}
   \rho=\left(
      \begin{array}{rrrr}
      u_1&&&\\
      &x_1&w&\\
      &w^{*}&x_2&\\
      &&&u_2
      \end{array}
   \right),
\end{equation}
the equation \cite{CW02}
\begin{equation}
   \label{eq:Conc}
   \tilde C=2\max\{|w|-\sqrt{u_1u_2}, 0\}
\end{equation}
serves for calculation of concurrence.
The expression (\ref{eq:Conc}) is a particular case of
Hill-Wootters formula \cite{HW97,W98} which allows to
calculate the pairwise concurrence in general two-qubit system.

Our density matrix (\ref{eq:rho-abvd}) has the form of
Eq.~(\ref{eq:rhobd}).
Using Eqs.~(\ref{eq:abdv}) we find that
\begin{equation}
   \label{eq:ConcDD}
   |w|-\sqrt{u_1u_2} = -v - \sqrt{ad}\le0 ,
\end{equation}
if $\Delta\leq-1$.
This inequality is valid for arbitrary external field $B$.
Hence, the concurrence
(\ref{eq:Conc}) and together with it the entanglement
(\ref{eq:EC}) are equal identically to zero.
So, in the dipole-dipole ($\Delta=-2$) dimer under question, the
quantum entanglement is absent for all temperatures and arbitrary
longitudinal fields.

\section{Information correlations}
\label{sec:QD}

In this section, we give a calculation of information correlations
in dipolar dimer both in and out magnetic field. 

\subsection{Arbitrary external field}
\label{sec:QDh}

Discord in a system with the density matrix having the structure
(\ref{eq:rho-abvd}) equals \cite{FWBAC10,LWF11}
\begin{equation}
   \label{eq:Q12}
   Q=\min\{Q_1, Q_2\} ,
\end{equation}
where
\begin{eqnarray}
   \label{eq:Q1}
  && Q_1 = S_A -S_{AB}
   - a\,{\rm log}_2\left(\frac{a}{a+b}\right) \nonumber \\ &&
   - b\,{\rm log}_2\left(\frac{b}{a+b}\right) 
   - b\,{\rm log}_2\left(\frac{b}{b+d}\right)\nonumber \\ &&
   - d\,{\rm log}_2\left(\frac{d}{b+d}\right) ,
\end{eqnarray}
\begin{equation}
   \label{eq:Q2}
   Q_2 = S_A -S_{AB}
   - \delta_1\,{\rm log}_2\delta_1
   - \delta_2\,{\rm log}_2\delta_2 ,
\end{equation}
and
\begin{equation}
   \label{eq:d12}
   \delta_{1,2}={1\over2}\,[1\pm((a-d)^2 + 4v^2)^{1/2}] .
\end{equation}
In equations (\ref{eq:Q1}) and (\ref{eq:Q2}), 
\begin{equation}
   \label{eq:SA}
   S_A = - (a+b)\,{\rm log}_2(a+b) - (b+d)\,{\rm log}_2(b+d)
\end{equation}
is the von Neumann entropy of reduced density matrix $\rho_A$
and
\begin{eqnarray}
   \label{eq:SAB}
  && S_{AB} = - a\,{\rm log}_2a - d\,{\rm log}_2d
            - (b+v)\,{\rm log}_2(b+v) \nonumber \\ &&
            - (b-v)\,{\rm log}_2(b-v)
\end{eqnarray}
equals the von Neumann entropy for the density matrix $\rho$
of full system.

The total correlation is $I=2S_A - S_{AB}$, and classical one
$C=I-Q$.
These relations and also (\ref{eq:Q12})--(\ref{eq:SAB}) together
 with Eqs.~(\ref{eq:abdv}) and (\ref{eq:Z}) define the quantum
discord, classical, and total correlations as functions of
the temperature and external magnetic field by arbitrary values
of parameters $D$ and $\Delta$. 

At high temperatures,
\begin{equation}
   \label{eq:Q1as}
   Q_1=\frac{1}{4\ln2}\left(\frac{D}{k_BT}\right)^{\!2}
   +\frac{\Delta}{8\ln2}\left(\frac{D}{k_BT}\right)^{\!3} + O(1/T^4) ,
\end{equation}
\begin{equation}
   \label{eq:Q2as}
   Q_2=\frac{1}{8\ln2}\,(1+\Delta^2)\left(\frac{D}{k_BT}\right)^{\!2}
   +O(1/T^3) . 
\end{equation}
Thus, the main terms of these expressions do not depend on the 
external magnetic field.
From the relations (\ref{eq:Q1as}) and (\ref{eq:Q2as}), one can see
that by $|\Delta|>1$ the discord, according to Eq.~(\ref{eq:Q12}),
is defined by the branch $Q_1$.
Hence, when
$\Delta>1$ or $\Delta<-1$, the quantum discord by high temperatures
behaves as
\begin{equation}
   \label{eq:QT8}
   Q\vert_{T\to\infty}\approx
   \frac{1}{4\ln2}\frac{1}{(k_BT/D)^2} .
\end{equation}
Thus, it does not depend on both external magnetic field or
interaction anisotropy $\Delta$.
With increasing the temperature, the quantum correlations decrease
according to the law $1/T^2$, i.~e., essentially rapidly than ordinary
statistical correlations which, how it can was emphasized above,
tend to zero according to the law $1/T$.

From equations presented above, we establish also that by high
temperatures the classical correlation is
\begin{equation}
   \label{eq:CT8}
   C\vert_{T\to\infty}\approx
   \frac{\Delta^2}{8\ln2}\frac{1}{(k_BT/D)^2} 
\end{equation}
and does not depend on external field.

\subsection{Dimer in absence of external field}
\label{sec:QDh0}

In important particular case of zero external field, the information
correlations can be calculated via the simpler formulas \cite{Luo08}.
In absence of field, $a=d$ and $S_A=S_B=1$.
Besides, using expressions (\ref{eq:Gzz}) and (\ref{eq:Gperp}),
it is not difficult to establish that by
$\Delta<-1$ one has 
$G_\parallel\geq|G_\perp|$ (below this will be seen in graphics).
Therefore, in absence of magnetic field, the classical part of
mutual correlations is
\begin{equation}
   \label{eq:C-h0}
   C = {1\over2}\,[(1+G_\parallel|){\rm log}_2(1+G_\parallel|)
   + (1-G_\parallel|){\rm log}_2(1-G_\parallel|) ,
\end{equation}
and the quantum discord equals
\begin{eqnarray}
   \label{eq:Qh0}
   &&Q={1\over4}\,[(1+2G_\perp -G_\parallel)
   {\rm log}_2(1+2G_\perp -G_\parallel)  \nonumber \\ &&
   - 2(1-G_\parallel) {\rm log}_2(1-G_\parallel)
   +(1-2G_\perp -G_\parallel) \cdot  \nonumber \\ &&
  \quad {\rm log}_2(1-2G_\perp -G_\parallel)] . 
\end{eqnarray}
How can one see from Eq.~(\ref{eq:C-h0}), the classical
correlation is completely determined by the longitudinal correlator
$G_\parallel$. 
Using expressions for the correlation functions (\ref{eq:Gzz}),
(\ref{eq:Gperp}) and setting $h=0$ in them, we get the following
formula for the discord,
\begin{eqnarray}
   \label{eq:QTh0}
  && Q(T) =\frac{\frac{D}{k_BT}{\rm sh}\frac{D}{k_BT}
   - {\rm ch}\frac{D}{k_BT}\ln[{\rm ch}\frac{D}{k_BT}]}
   {({\rm ch}(\frac{D}{k_BT}) + \exp(-\frac{D\Delta}{k_BT}))\ln (2)} .\nonumber \\ &
\end{eqnarray}

\begin{figure}[t]
\begin{center}
\epsfig{file=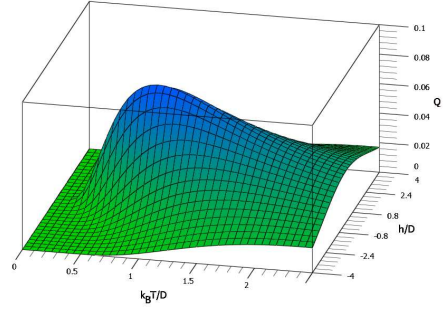,width=8cm}
\caption{
Quantum discord in dipolar system ($\Delta=-2$) as a function of 
temperature and external longitudinal magnetic field.}
\label{fig:fig1}
\end{center}
\end{figure}

For a dipolar dimer ($D>0$, $\Delta=-2$) in zero field and at lower
temperatures, we have
\begin{equation}
   \label{eq:QT0}
   Q\vert_{T\to0}\approx
   \frac{1}{2}\exp{(-D/k_BT)} .
\end{equation}
Arbitrary order derivatives with respect to the temperature for the
function in the right hand of this equation is zero at $T=0$.
Therefore, by small deviation of temperature from absolute zero, 
$Q\approx0$.
This is connected with existence of a gap in the energy spectrum
of the system.

\begin{figure}[t]
\begin{center}
\epsfig{file=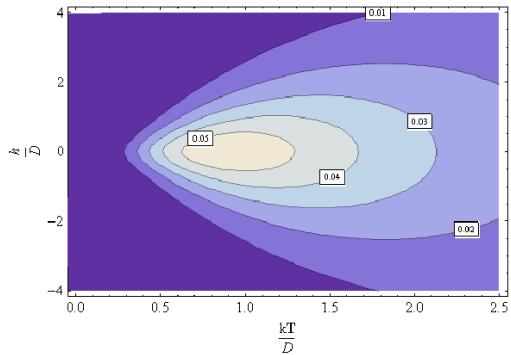,width=7cm}
\caption{
Isolines of quantum discord in dipolar system ($\Delta=-2$).}
\label{fig:fig2}
\end{center}
\end{figure}

\section{Discussion}
\label{sec:Discuss}

A behaviour of quantum discord in the dipole-dipole system under
question ($\Delta=-2$) is shown in Fig.~\ref{fig:fig1}.
We see the smooth hill-like surface stretched in the temperature
axis direction. 
Along a straight line $h=0$, two ridges go; one ridge goes in the
direction of temperature decreasing and the other in the direction
of its increasing. 
The surface is symmetrical under reflection in vertical plane
going through the hill top and the straight line
$h=0$ in the temperature-field plane.
By given temperature, the discord is maximal in absence of external
magnetic field, that is, the field leads only to a suppression
of quantum correlation.
At the absolute zero temperature, the discord in dimer is
identically equal to zero.
In the high temperature limit, quantum correlation is also
vanished. 
In Fig.~2, the cross sections (profiles) of a discord surface are
shown by different values of $Q$.
It is seen that the isolines form a set of non-crossing ovals.

\begin{figure}[t]
\begin{center}
\epsfig{file=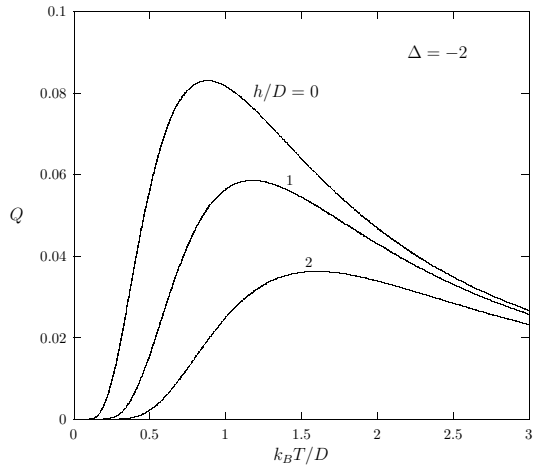,width=7cm}
\caption{
Discord isoterms by different values of external field.}
\end{center}
\label{fig:fig3}
\end{figure}

To study in detail the behavior of discord, we represented in
\ref{fig:fig3} its temperature dependence by different values of external field. 
The discord reaches the largest value at
$h=0$ and
\begin{equation}
   \label{eq:Tm}
   k_BT_m/D = 0.881\,297\ldots\ . 
\end{equation}
The discord in this point equals 
\begin{equation}
   \label{eq:Qm}
   Q_m = 0.083\,061\ldots\ .
\end{equation}
what is $8.3\%$ of maximal value which is possible in any two-qubit
system.
The value (\ref{eq:Tm}), we have found both by a numerical search of
maximum for the function $Q$ of $k_BT/D$ and from solution of
transcendental equation
\begin{equation}
   \label{eq:Tm-eq}
   x(e^{\Delta\cdot x} + {\rm ch}x + \Delta\cdot{\rm sh}x)
   =({\rm sh}x + \Delta\cdot{\rm ch}x)\ln({\rm ch}x),
\end{equation}
where $x=D/k_BT$.
This equation follows from the condition $\partial Q/\partial T =0$
in which the function $Q(T)$ is given by expression (\ref{eq:QTh0}). 

One can see from Fig.~3 that, with increasing the field, the discord
maximum is moved in the direction of higher temperatures.
This is a good feature.
However by this the value of discord at the maximum is less than its
value at the same temperature in absence of external field.

We see also that with the help of external field and temperature
(these parameters are in our hands),
one can control the discord varying its value from zero to 0.083~bit
per a dimer. 

\begin{figure}[t]
\begin{center}
\epsfig{file=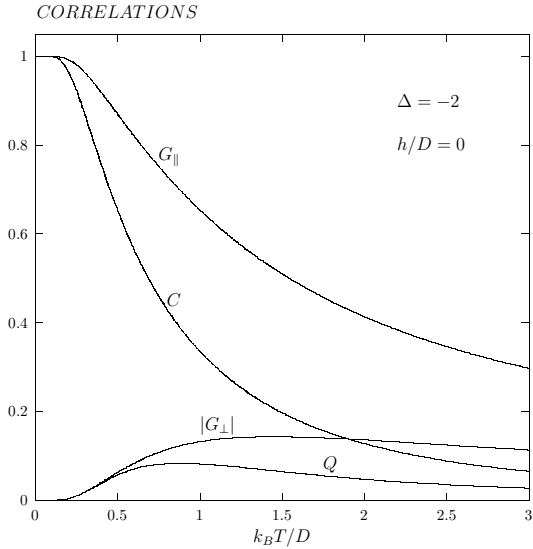,width=7cm}
\caption{
Temperature behavior of statistical ($G_\parallel$ and $|G_\perp|$),
classical ($C$), and quantum ($Q$) correlations in absence of magnetic
field.}
\label{fig:fig4}
\end{center}
\end{figure}

Let us consider the case when the dipolar dimer is in absence of
external field.
The temperature dependences of ordinary spin-spin correlation functions
$G_\parallel$ and $|G_\perp|$ ($G_\perp\leq0$, therefore its 
absolute value is taken), as well as total classical correlation $C$
and discord $Q$ are presented in Fig.~4.
At zero temperature, both longitudinal and total classical
correlation are equal to their maximal magnitudes.
The temperature increasing leads only to their decrease.
Both the value of transverse correlation $|G_\perp|$ and quantum discord
are zero at $T=0$.

To clarify the situation in correlation behavior at zero, we consider the
limit of density matrix (\ref{eq:rho-abvd}) when $T\to0$.
Using the expressions for its matrix elements (\ref{eq:abdv}) one finds 
\begin{equation}
   \label{eq:rho-h0t0}
   \rho\vert_{T=0}=\left(
      \begin{array}{rrrr}
      1/2&&&\\
      &0&&\\
      &&0&\\
      &&&1/2
      \end{array}
   \right) .
\end{equation}
This matrix has the expansion
\begin{equation}
   \label{eq:rho-cl}
   \rho={1\over2}(|0\rangle\langle0|_A\otimes|0\rangle\langle0|_B
   + |1\rangle\langle1|_A\otimes|1\rangle\langle1|_B ,
\end{equation}
i.~e., it is a sum of the direct products of separate particles. 
Such state belongs to the class of purely classical ones
\cite{PSVW10} and therefore, according to the criteria  
\cite{DVB10}, all quantum correlations in it are entirely absent.
This explains that we obtained $Q\vert_{T=0}=0$.
On the other hand, classical correlations are available and equal
the total correlations $C=I$.

So the behavior of system at  $T=0$ is purely classical.
However, with increasing the temperature, the behavior
acquires the quantum features, i.~e.,
the temperature leads to the generation of quantum correlations. 
Such unusual phenomenon (it is much ordinary when the temperature
destroys the quantum states) one can explain by following.
In the Hamiltonian (\ref{eq:Hxxz}) with 
$\Delta<-1$, the purely classical ferromagnetic contribution
\begin{equation}
   \label{eq:Hzz}
   H_{zz}=\Delta D\sigma_1^z\sigma_2^z 
\end{equation}
is dominate at lower temperatures.
The transverse components of spins are actually ``frozen''
(the value $G_\perp=0$ witnesses this).
With increasing the temperature these degrees of freedom
``revive'' and the system from classical becomes a quantum one. 
But when the temperature fluctuations begin to exceed the system
energy gap, the temperature produces its usual destroying action
and the correlations go down to non. 

\begin{figure}[t]
\begin{center}
\epsfig{file=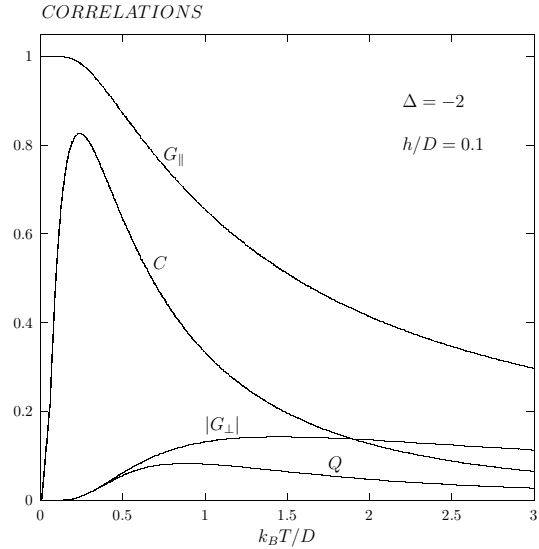,width=7cm}
\caption{
The same as in Fig.~4 but in presence of a field.}
\label{fig:fig5}
\end{center}
\end{figure}

From Fig.~4 and asymptotical behavior of correlations, it is not
difficult to notice a correspondence in qualitative behavior,
on the one hand,  $C$ and $G_\parallel^2$ and, on the other hand,
$Q$ and $G_\perp^2$.
The latter correspondence is not an accident.
Indeed, if the quantum correlation is measured by geometrical
discord $Q_g$, i.~e., by a distance (in sense of the Hilbert-Schmidt
norm) from the given state $\rho$ to the nearest classical one,
the general formula \cite{DVB10,DLMR12} applied to 
the density matrix (\ref{eq:rhoB}) by an absence of external field
yields 
\begin{equation}
   \label{eq:Qg}
   Q_g=G_\perp^2 .
\end{equation}

Consider now the situation in a presence of external magnetic
field.
In Fig.~5, the behavior of different correlations is again shown
but for $h/D=0.1$.
It is seen that correlations $G_\parallel$, $G_\perp$, and $Q$
were changed weakly.
But the behavior of classical correlation $C$ undergoes an
essential change --- it tends now to zero when $T\to0$.
At the point of absolute zero temperature, there are no any
correlations in the system at all. 

\begin{figure}[t]
\begin{center}
\epsfig{file=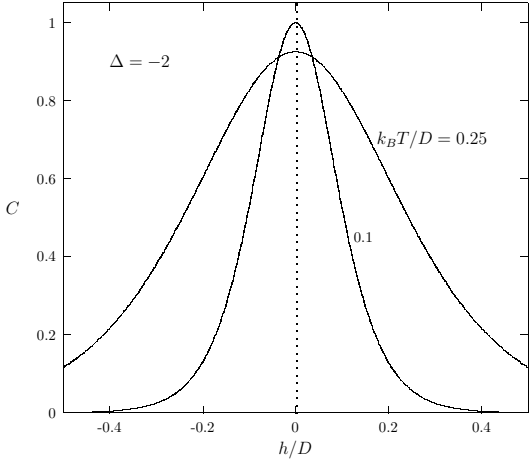,width=7cm}
\caption{
Classical correlation vesus the field by diffrrent values
of temperature.}
\label{fig:fig6}
\end{center}
\end{figure}

To understand a happening, we turn out again to the density
matrix.
The matrix (\ref{eq:rho-abvd}) in the limit $T\to0$ is equal now
to both
\begin{equation}
   \label{eq:rho-h1t0}
   \rho\vert_{T=0}=\left(
      \begin{array}{rrrr}
      1&&&\\
      &0&&\\
      &&0&\\
      &&&0
      \end{array}
   \right) , 
\end{equation}
if $h>0$ or 
\begin{equation}
   \label{eq:rho-h1mt0}
   \rho\vert_{T=0}=\left(
      \begin{array}{rrrr}
      0&&&\\
      &0&&\\
      &&0&\\
      &&&1
      \end{array}
   \right) , 
\end{equation}
if $h<0$.
These states correspond to the completely orderings of spins along
the field.
Both matrices (\ref{eq:rho-h1t0}) and (\ref{eq:rho-h1mt0}) are
factorized into a direct product of two density matrices $2\times2$,
\begin{eqnarray}
   \label{eq:rho-dp}
  && \left(
      \begin{array}{rrrr}
      1&&&\\
      &0&&\\
      &&0&\\
      &&&0
      \end{array}
   \right)=
   \left(
      \begin{array}{rr}
      1&0\\
      0&0
      \end{array}
   \right)\otimes
   \left(
      \begin{array}{rr}
      1&0\\
      0&0
      \end{array}
   \right) ,\nonumber \\ &&
   \left(
      \begin{array}{rrrr}
      0&&&\\
      &0&&\\
      &&0&\\
      &&&1
      \end{array}
   \right)=
   \left(
      \begin{array}{rr}
      0&0\\
      0&1
      \end{array}
   \right)\otimes
   \left(
      \begin{array}{rr}
      0&0\\
      0&1
      \end{array}
   \right) . 
\end{eqnarray}
This means that the states (\ref{eq:rho-h1t0}) and
(\ref{eq:rho-h1mt0}) are completely uncorrelated and therefore
any correlations are absent in them.

Emphasize that $G_\parallel\neq0$ at $T=0$ does not
contradict to the said above.
In the factorized state must be zero only the centered correlator.
By $h\neq0$, the spins at the point of absolute zero temperature
are ordered and therefore 
\begin{equation}
   \label{eq:ccor}
   \langle(\sigma_1^z -\langle\sigma_1^z\rangle)
   (\sigma_2^z -\langle\sigma_2^z\rangle)\rangle=0 .
\end{equation}
Thus, any statistical relation between spins is also absent.

Fig.~6 shows the behavior of classical correlation $C$ upon
the external field.
The curves have the form of bell-like splashes. 
Their largest value is arrived at in the point when the field
vanishes.
With decreasing the temperature, the maximums become more narrow
and their value tends to the value $C=1$.
At $T=0$, a splash becomes infinitely thin, 
\begin{equation}
   \label{eq:CT0}
   C=\begin{cases} 1,& h=0\\      0,&  h\neq 0  \end{cases} .
\end{equation}
The external field sweeping will on a moment lead to a spasmodic jump
appearance of classical correlation and at the same time to a
disappearance of this correlation by going the field of the zero point. 
By this, the quantum correlation in the system does not arise.

Let's turn to the available experimental data.
Measurements performed by the nuclear magnetic resonance (NMR) at room
temperature show that in gypsum crystals ${\rm CaSO_4\cdot2H_2O}$
the distance between protons in each water molecule is $r=0.158$~nm
\cite{P48} (see also, e.~g., the book \cite{A61}).
For protons, the gyromagnetic ratio, as it is known \cite{G80},
equals $\gamma=2.675\cdot10^8$~rad/(c$\cdot$T),
therefore the dipole-dipole coupling constant (\ref{eq:D}) in gypsum
is $D/k_B=0.73~\mu$K (in temperature units).
Consequently, in accord with (\ref{eq:Tm}), the maximum discord
value $Q_{max}=0.083$ will arrive at the temperature 0.64~$\mu$K.
At room temperature ($T=300$),
the discord in gypsum, according to
(\ref{eq:QT8}), must be equal to $Q\sim2\cdot10^{-18}$.
In spite of extremely small value of quantum correlations in
spin-nuclear systems at room temperatures, at present the attempts are
undertaken to detect their by NMR methods \cite{AMC11,MACSP12,PMTL11,KRMP12}. 

As an another example, let us consider 1,2-dichloroethane 
${\rm ClH_2C-CH_2Cl}$.
In this compound, two protons at each carbon atoms
are coupled much stronger between themselves by dipole-dipole
interaction than with protons belong to an other carbon atom.
NMR measurements performed on solid dichloroethane at the
temperature 90~K have shown that here $r=0.17(2)$~nm \cite{GKPP49}
(see also \cite{A61,G80}). 
Using again the relations (\ref{eq:D}) and (\ref{eq:Tm}),
we estimate the temperature for the discord maximum in
this substance as $T_m=0.517~\mu$K.
At the temperature 90~K, the value of quantum correlations must equal
$Q\sim1.5\cdot10^{-17}$.

\section{Conclusions}
\label{sec:Concl}

In the paper, a study of information correlations in dipolar dimers
both in absence of external magnetic field and in the field directed
along the longitudinal axis of a dimer has been performed.
It bas been shown that the quantum correlations are completely absent
at the absolute zero temperature and arbitrary strength of
magnetic field but they arise with increasing the temperature.
We proposed an interpretation such of phenomenon.

In the high temperature region, the discord obeys the law 
$$
Q\sim G_\perp^2\sim1/(k_BT/D)^2 .
$$
This, in particular, means that at room, nitrogen, or helium
temperatures, the non-zero quantities of spin-spin correlations
between the $xx$ or $yy$ spin components can serve as a witness
of quantum correlations.
The spin-spin correlations go down with increasing the temperature
according to the essentially slow law $T^{-1}$ and one can directly
measure them, for example, in scattering experiments.
 
It was shown also that the classical correlations have a sharp
maximum in the point of zero external magnetic field when $T\to0$. 

A qualitative analogy between the quantum discord and the squared
correlator $G_\perp$ has been found. 
In supporting this observation, it was shown that the geometrical
discord equals $Q_g=G_\perp^2$.

In the paper, the estimates for the thermal quantum discord between
spins of hydrogen nucleous ${}^1$H in gypsum and 1,2-dichloroethane
have been done. 

Our investigation can be extended to many-nuclear clusters if to
perform a density matrix reduction for all spins except any two.
Experimental NMR data are available for clusters in a form, for
example, triangles, tetrahedron, linear magnetic structures, etc.,
\cite{A61,AS61,ZDB09}.

\section*{ACKNOWLEDGMENT}

This research was supported by the program No.~8 of
the Presidium of RAS.




\newpage

\end{document}